\documentstyle[prl,aps,epsf]{revtex}
\begin{document}
\draft
\title{Fractional and unquantized dc voltage generation in
THz-driven semiconductor superlattices}
\author{
K. N. Alekseev$^{1,2}$\thanks{E-mail: Kirill.Alekseev@oulu.fi},
E. H. Cannon$^3$, F. V. Kusmartsev$^{4}$, and D. K. Campbell$^5$ \\
$^1$Division of Theoretical Physics, Department of Physical Sciences,\\
Box 3000, University of Oulu FIN-90014, Finland \\
$^2$Theory of Nonlinear Processes Laboratory, \\
Kirensky Institute of Physics, Krasnoyarsk 660036, Russia\\
$^3$Department of Electrical Engineering,\\
University of Notre Dame, Notre Dame, IN 46556, USA\\
$^4$Department of Physics, Loughborough University, Loughborough
LE11 3TU, UK\\
$^5$Departments of Electrical and Computer Engineering and Physics,\\
Boston University, Boston, MA 02215, USA}
\maketitle
\begin{abstract}
We consider the spontaneous creation of a dc voltage across a strongly coupled
semiconductor superlattice subjected to THz radiation. We show that the dc
voltage may be approximately proportional either to an integer or to a
half-integer multiple of the frequency of the applied ac field,
depending on the ratio of the characteristic scattering rates of conducting
electrons. For the case of an ac field frequency less than the characteristic
scattering rates, we demonstrate the generation of an unquantized dc voltage.
\end{abstract}
\pacs{PACS numbers: 73.21.Cd; 72.20.Ht; 05.45.-a}
The theoretical analysis of nonlinear transport properties of strongly-coupled
semiconductor superlattices (SSLs) irradiated by a high-frequency electric
field began already in the mid 1970's \cite{bass}.  Recently, experimental
progress in creating powerful sources of THz radiation, the development of
a coupling technique \cite{exp-thz,winnerl97}, and improvement in the
fabrication technology of microstructures leading to very high carrier mobility
\cite{exp-samples} have stimulated many new theoretical
investigations of this long-studied problem. Among these
works have been studies of strongly nonlinear effects
including multistability \cite{multistab}, short pulse generation
\cite{zharov00}, chaos \cite{alekseev96,cao00,yevtushenko}, and
spontaneous generation of a dc voltage in a purely ac-driven SSL
\cite{ignatov95,alekseev98,alekseev99,romanov00}.
\par
In this paper we investigate the last effect in further detail and
discuss the appearance of new dc voltage states, which are a
generalization of the integer dc voltage states in SSLs previously
described \cite{ignatov95,alekseev98}. These states are related to the complex
dynamics of miniband electrons in an SSL and the formation of the
Wannier-Stark ladder in a purely ac driven SSL. Two distinct mechanisms are
known for the spontaneous generation of a dc bias
in purely ac-driven SSLs \cite{ignatov95,alekseev98,romanov00}. Both these
nonlinear mechanisms work in SSLs with a high mobility and a relatively high
level of doping, when the effects of a self-consistent electric field
generated by an electron's motion become significant.
The first mechanism arises if the ac field frequency, $\omega$, is
much greater than the plasma frequency, $\omega_{pl}$, and is related
\cite{ignatov95,romanov00} to an instability caused by absolute negative
conductivity in the ac-driven SSL \cite{anc}. The effect has been attributed
\cite{ignatov95} to the phenomena of dynamical localization of electrons
\cite{dunlap86} and miniband collapse in a collisionless SSL \cite{holthaus}.
It was reported that the generated dc bias is such that the ``induced Bloch
frequency'' $\omega_B=e a E_{dc}/\hbar$
($E_{dc}$ is the spontaneously generated dc electric field and $a$
is the SSL period) is approximately equal to the ac field frequency $\omega$
\cite{ignatov95,romanov00,dunlap93}.
\par
The second mechanism is responsible for dc bias generation when the
ac field frequency is near the plasma resonance, $\omega\simeq\omega_{pl}$;
it arises for a smaller ac field strength than is required in the previous
situation \cite{alekseev98}. The instability responsible for the dc bias
generation in SSLs with strong enough electron scattering also results in
chaotic motion in the case of small scattering rates or for a
collisionless SSL \cite{alekseev96,alekseev98}.  The creation
of a dc bias may be qualitatively explained \cite{alekseev99} and classified
\cite{romanov00} using the semiclassical theory of wave-mixing in SSLs
\cite{wave-mix}.
\par
In this paper we re-examine the problem of spontaneous dc voltage
generation in an SSL subjected to a THz electric field.  We show that,
depending on the relative values of the scattering rates and the
ac field frequency, a variety of different dc voltage states can
exist, including both integer and half-integer quantized states, for
which the induced Bloch frequency is approximately an integer or
half-integer multiple of the ac field frequency, and completely
unquantized states. In particular, if the electron velocity relaxation
rate, $\gamma_v$, is sufficiently different from the electron energy
relaxation rate, $\gamma_\varepsilon$, and $\omega\gtrsim\gamma_v$,
we find integer states with $\omega_B\approx n\omega$ ($n=\pm 1,\pm
2,\ldots$); while for $\gamma_v=\gamma_\varepsilon$,
the states are close to the half-integer states $\omega_B\approx n\omega/2$.
In contrast, in the case of low-frequency driving or high damping,
$\omega<(\gamma_v,\gamma_\varepsilon)$, the dc voltage states are unquantized.
\par
We study electron transport through a single miniband, spatially homogeneous
SSL with period $a$ and miniband width $\Delta$, which is subjected
to an ac electric field $E(t)=E_0\cos\omega t$ along the SSL axis.
For the tight-binding energy-quasimomentum
dispersion relation $\varepsilon(k)=(\Delta/2)\left[1-\cos(ka)\right]$
($k$ is the electron wave vector along the axis of SSL), the dynamics of
electrons is described by the superlattice balance equations
\cite{ignatov95,alekseev98,romanov00}
\begin{eqnarray}
\dot{v}&=&u w -\gamma_v v,\nonumber\\
\dot{w}&=&-u v -\gamma_\varepsilon (w-w_{eq}),\label{balance}\\
\dot{u}&=&\omega_{pl}^2 v-\alpha u+I_{ext}(t),\nonumber
\end{eqnarray}
where $v=m_0 \overline{V} a/\hbar$,
$w=(\overline{\varepsilon}-\Delta/2)
(\Delta/2)^{-1}$ and $w_{eq}$ are a scaled electron velocity, a scaled
electron energy, and an equilibrium value of scaled electron energy,
respectively, and $m_0=(2\hbar^2)/(\Delta a^2)$ is the effective mass at the
bottom of miniband. The scaled variables $v(t)$ and $w(t)$ are proportional to
the variables $\overline{V}(t)$ and $\overline{\varepsilon}(t)$, which are
the electron velocity and energy averaged over the time-dependent distribution
function satisfying the Boltzmann equation. The lower (upper) edge
of the miniband corresponds to $w=-1$ ($w=+1$), and the value of $w_{eq}$ is a
function of the lattice temperature (for a thermal equilibrium $w_{eq}<0$).
The variable $u(t)$ is related to the electric field inside the SSL $E(t)$
as $u=e a E/\hbar$. In deriving Eqs. (\ref{balance}),
we assumed that the electrical properties of an SSL of total length
$l$ and cross-section $S$ can be modeled by an equivalent high-quality circuit
\cite{ignatov95} which consists of a capacitor $C=(\epsilon_0 S)/(4\pi l)$
($\epsilon_0$ is the average dielectric constant for the SSL) driven by an ac
current of the form $I_{ext}=-\omega_s\omega\sin\omega t$, where
$\omega_s=e E_0a/\hbar$, in our scaled units.
\par
The first Eq. of set (\ref{balance}) describes an acceleration of
electrons under the action of electric field $E(t)$ and their slowing
down caused by an effective friction due to scattering. In original
dimensional variables, the term $u w$ in r.h.s. of the first Eq. looks like
$e E/m(\overline{\varepsilon})$, where $m(\overline{\varepsilon})=
m_0/(1-2\overline{\varepsilon}/\Delta)$ is energy-dependent effective mass
of the electrons in SSL's miniband. The second Eq.
describes a balance of electron's energy gain under the action of
electric field and energy loss due to scattering. Finally, the third
equation describes a balance of diffusive, external and displacement
currents in the SSL. The degree of nonlinearity in Eqs. (\ref{balance})
is controlled by the value of miniband plasma frequency,
$\omega_{pl}=\left( 4\pi e^2 N/m_0\epsilon_0\right)^{1/2}$,
which is a function of the electron doping density $N$,
while the parameter $\alpha$ determines the quality of the effective
circuit ($\alpha\ll\omega_{pl},\omega$).
\par
The relaxation processes for miniband electrons are characterized by
an average energy scattering rate, $\gamma_\varepsilon$,
as well as by an average velocity scattering rate,
$\gamma_v=\gamma_\varepsilon+\gamma_{el}$, where  $\gamma_{el}$ is an average
rate of elastic collisions \cite{winnerl97,ignatov95}.
The scattering rates, $\gamma_v$ and $\gamma_\varepsilon$, can have
different values depending on the material, the doping density, the
temperature, etc.  In particular, for microstructures with modulation
doping \cite{modulation-doping}, the ionized
impurities are spatially separated from electrons, which greatly reduces
the elastic scattering rate $\gamma_{el}$ so that
$\gamma_v\approx\gamma_\varepsilon$. In contrast, for many vertical SSLs
operating at room temperature, the scattering rate for electron velocity is
about of order of magnitude greater than the characteristic scattering rate of
electron energy, $\gamma_v/\gamma_\varepsilon\approx 10$
\cite{winnerl97,ignatov95}.
\par
We solve the nonlinear balance Eqs. (\ref{balance}) numerically for the
initial conditions $v(0)=0$, $w(0)=w_{eq}=-1$, with the circuit damping
rate $\alpha/\omega_{pl}=0.01$, and for two typical sets of relaxation
constants: (i) $\gamma_v/\gamma_\varepsilon=10$,
$\gamma_\varepsilon/\omega_{pl}=0.01$, and (ii)
$\gamma_v/\gamma_\varepsilon =1$, $\gamma_\varepsilon/\omega_{pl}=0.1$.
After removing the transients, we calculate the time-average value
$\langle u\rangle$, which gives the value of the Bloch frequency determined
by the spontaneously generated dc bias $E_{dc}$:
$\omega_B\equiv eaE_{dc}/\hbar=\langle u\rangle$.  Figs. 1 and 2 present the
results of computations of $\langle u\rangle$ for 201 value of
$\omega_s$ equally distributed in the range
$0\le\omega_s\le 2\omega_{pl}$ for each driving frequency.
\par
For the first set of relaxation rates and for $\omega>2\gamma_v$,
plateaus of integer-quantized states are clearly observable:
$\omega_B\approx n\omega$, with $n=\pm 1,2,3,4$ in Fig.1.
However, at low frequencies, $\omega< 2\gamma_v$,
instead of steps there is a region of unquantized dc states.
The dependence of the induced Bloch frequency $\langle u\rangle$ on
the ac field frequency $\omega$ for the second set of damping parameters is
presented in Fig.2. Here, qualitative differences from Fig. 1 appear,
including 1) the existence of half-integer states, specifically states
with $n\approx\pm 1/2$, and 2) the nonzero width of the $n=0$ plateau.
One can expect the width of a plateau to be equal to
the scattering rate, $0.1\omega_{pl}$ in this case.
As a result, we have plotted the lines
$n\pm 0.1\omega_{pl}/\omega$ for $n=0,\pm 1, \pm 2,\pm 3,-4,-5$ in Fig.2.
We found that, indeed, most points for a given plateau lie in the region
demarcated by the two lines with same $n$. In contrast to Fig. 1, there are
states with $\langle u\rangle/\omega<0.1$, but $\langle u\rangle\not=0$.
Also, some points fall very near the line $\langle u\rangle=\pm 0.5\omega$.
As an example we refer to the solution of Eqs. (\ref{balance}) for
$\omega/\omega_{pl}=0.6$ and $\omega_s/\omega_{pl}=0.6$ (other parameters are
same as in Fig.2), which in the phase space
corresponds to a symmetry-broken limit cycle\footnote{
This is a limit cycle whose projection on the $v-u$ plane is not
symmetric about the origin, in contrast to a symmetric limit cycle
(see \cite{alekseev98}).}
with $\langle u\rangle=0.289\omega_{pl}$.
\par
If $\omega/\gamma$ is less or order of unity  but
$I_{ext}/\omega^2_{pl}$ is large enough
and $\gamma_v=\gamma_\varepsilon\equiv\gamma$, then
the existence of nonquantized and half-integer dc voltage states can be
demonstrated analytically. In these limits, the motion on an attractor
of dynamical system (\ref{balance}) is governed by the following
pendulum equation \cite{alekseev00}
\begin{equation}
\label{pend}
\ddot{\theta}+\gamma\dot{\theta}+(-w_{eq}) \omega_{pl}^2
\sin\theta=-2 I_{ext}(t),
\end{equation}
where we made the substitutions: $v=(w_{eq}/2)\sin\theta$,
$w=(w_{eq}/2)(1+\cos\theta)$ and
\begin{equation}
\label{field}
u=\dot{\theta}/2+\gamma\tan(\theta/2).
\end{equation}
The dc voltage $\langle u\rangle$ can be obtained as a result
of averaging over time of r.h.s. of Eq. (\ref{field}).
The equation (\ref{pend}) looks same as a motion
equation of the well-known Stewart-McCumber model from the theory of
ac-driven Josephson juctions \cite{kautz}. However, the principal
difference from this model also exists. In the Stewart-McCumber model
the voltage across the Josephson junction is proportional to the velocity of
pendulum \cite{kautz}, while in our case the voltage is a function of both
velocity and co-ordinate (see Eq. (\ref{field})). This difference plays a
principal role in the explanation of existence in our case of both unquantized
and only approximately quantized
dc voltage states, which are absent in the Josephson junction model.
\par
The damped and driven pendulum (Eq. (\ref{pend})) have two
distinct types of attractors with regular dynamics: rotating
and oscillating \cite{kautz,d'humieres}. Majority of rotating states
are phase-locked, {\it i.e.}, $\langle\dot{\theta}\rangle=(n/l)\omega$
($n$ and $l$ are integer numbers); they are observable mainly at
$\omega\gtrsim\gamma$ \cite{d'humieres}. As it is evident from
Eq. (\ref{field}), such phase-locked rotational pendulum states are
responsible for the generation of half-integer dc voltage states in SSL:
$\langle u\rangle\approx (n/2l)\omega$. Note that
these corrections to these quantized states arise from
the contribution of $\theta$-dependent term in Eq. (\ref{field}).
The corrections can give the dependence of $\langle u\rangle$ on ac
current amplitude, $\omega_s$, its frequency, $\omega$, and scattering
constant, $\gamma$. However, it can be shown that a relative contribution into
these corrections is controlled by the parameter
$\gamma/\omega_{pl}$ \cite{alekseev00}. Therefore,
the dependence of approximately quantized voltage on ac field
amplitude and frequency is weak until $\gamma/\omega_{pl}\ll 1$.
\par
For oscillating attractors, the equality
$\langle\dot{\theta}\rangle=0$ is always valid
\cite{d'humieres}. If additionally stationary pendulum
oscillations are symmetric, {\it i.e.} $\langle\theta\rangle=0$,
then the dc voltage can not be generated, $\langle u\rangle=0$.
However, the pendulum (Eq. (\ref{pend})) can demonstrate
{\it symmtry-broken oscillations}, for which $\langle\theta\rangle\neq 0$
\cite{d'humieres,sb-def}. The symmetry-broken oscillations mainly
exist at the low frequencies of external force; they can survive
at $\omega<\gamma$ even for a strong damping, when
$\gamma/\omega_{pl}\gtrsim 1$ \cite{sb-strong}. As it is evident from
Eq. (\ref{field}), just the symmetry-broken oscillations corresponding to
$\langle\dot{\theta}\rangle=0$ and $\langle\theta\rangle\neq 0$, are
responsible for the generation of the unquantized dc voltage in SSL.
\par
Unquntized dc voltage states exist in the low frequency region for different
ratious of the scattering constants. In order to understand the transition from
quantized to unquantized dc bias states as the scattering rates increase while
maintaining $\gamma_v\gg\gamma_\varepsilon$, we present in Fig.3 the dependence
of $\langle u\rangle$  on $\omega$ for strong damping.  As it is evident
from this figure, the strong damping destroys all quantized states; dc bias
generation persists only for some unquantized states.
\par
It is instructive to consider the mechanism of unquantized dc bias
generation in the terms of energy levels structure.
We offer the following qualitative explanation of the underlying
physics: when the electron scattering rates are sufficiently small and
the amplitude of the ac field is large enough,
the SSL spontaneously creates a Wannier-Stark ladder with the spacing,
$\omega_B$, that makes multiphoton absorption of ac field most
effective, {\it i.e.} $\omega_B\approx n\omega$. In the pendulum analogy this
 means the appearance of the pendulum rotations, which frequency of rotation
is proportional to the generated dc bias.
However, for larger damping, the rotations are ceased and therefore
the symmetry broken oscillations remain, that means that only a small bias can
be generated in the SSL.
Hence the spacing of the Wannier-Stark ladder states is less than
the ac field frequency except for very low frequencies $\omega\lesssim\gamma$.
In this case, the broadening of the self-organized ladder levels is
quite comparable  with their spacing; it is therefore practically
impossible to achieve quantized values of the voltage (or a rotation of the
pendulum). In a real space picture such a situation should correspond to an
appearence of some kind of semi-localized Wannier-Stark wave functions;
this structure of the ladder is reminiscent of the wave function of
biased short superlattice \cite{helm,vagov}, where the spacing between energy
levels depends non-linearly on the bias voltage \cite{vagov}. In our case, a
dc bias is created by ac field, thus we have a weak dependence of ladder
spacing on ac field strength via the self-induced bias.
\par
We have performed a systematic numerical study of the positions and widths of
different plateaus and of the unquantized states for many values of
the driving amplitude and frequency and several different initial conditions at
different damping levels \cite{thesis}; the results are in a qualitative
agreement with situation described above.
We now make several remarks on the results of this search.
First, we found no indication of the noninteger (fractional)
dc states for $\gamma_v\ne\gamma_\varepsilon$. However, for
$\gamma_v=\gamma_\varepsilon$, we find that the $1/2$-dc-states are quite
common. Moreover, for weak enough damping, we additionally saw a few
dc states that are very close to fractional states
of the form $n/k$ with $n$ being an integer and $k$ always {\em being an even
integer}. Such dc states are formed by the symmetry-broken limit cycles with
large even periods. As examples, we refer to $3/2$-states formed by period-12
and period-24 limit cycles, as well as to the $7/6$-state formed by a period-12
limit cycle; both occur for
$\gamma_v=\gamma_\varepsilon=0.05\omega_{pl}$ and $\alpha/\omega_{pl}=0.01$.
We should note, however, that for weak damping, chaotic behaviour is quite
typical \cite{alekseev96,alekseev98} and that both the stable limit
cycles and the asymmetric chaotic attractors, which are responsible for the
generation of {\em stable quantized dc voltage states} in the
SSL, occupy only a small amount of parameter space of the system
\cite{thesis,alekseev00}.
\par
Importantly, it appears possible to achieve the parameter values
used in our simulations in the conditions of experiments.
To begin with we refer to the experiments with a heavily doped
SSL ($N=8\times 10^{16}$ ${\rm cm^{-3}}$) having at room
temperatures scattering constants $\gamma_v\approx 10^{13}$ ${\rm s^{-1}}$ and
$\gamma_\varepsilon\simeq 0.1\gamma_v$ \cite{winnerl97}.
Using the values $a=4.8$ nm, $\Delta\approx 50$ meV and
$\epsilon_0\approx 13$ \cite{winnerl97}, we get $\omega_{pl}=1.2\times 10^{13}$
rad/s and $\gamma/\omega_{pl}\approx 0.84$
(compare, {\it e.g.}, with data of Fig.3).
\par
Longer scattering time provide SSLs based on the cleaved
edge overgrown technique: $\gamma^{-1}\approx 3$ ps at low
temperatures \cite{exp-samples}. In conditions of experiment
\cite{exp-samples}, for $a=10$ nm, $\Delta\approx 20$ meV,
$N=N_s b\approx 3\times 10^{15}$ ${\rm cm^{-3}}$
($N_s=3\times 10^{11}$ ${\rm cm^{-2}}$ is an electron areal density
and $b\approx 10^{-4}$ cm is sample's thickness),
the miniband plasma frequency is $\omega_{pl}=3.1\times 10^{12}$
rad/s providing $\gamma/\omega_{pl}\approx0.1$
({\it cf.} with data of Fig. 2). In our numerical simulations
we used the ac field strengths satisfying $\omega_s\leq 2\omega_{pl}$,
what in the physical units correspond to the realistic values
$E_0\leq 5$ kV/cm. Necessary for the observation of our effects ac
field frequencies belong to the THz range.
Finally, we should also note that for all listed cases, the standard condition
of validity of the single-miniband approximation,
$\Delta\gg\hbar\omega_s$ \cite{anc}, is well satisfied because
$\omega_s\hbar/\Delta\lesssim 0.1$ .
\par
In summary, we have shown that a semiconductor superlattice irradiated by
a high-frequency electric field can spontaneously generate a dc bias,
which can be quantized in approximately integer or particular fractional
ratios of the driving frequency, or completely unquantized.
In this respect, the effects in semiconductor superlattices are
no less rich than their counterparts in Josephson junctions subjected
to a microwave field, where the exactly integer and the exactly fractional dc
voltage states (``phase-locked states'') are known \cite{kautz}.
\par
We thank Anatoly Ignatov, Pekka Pietil\"{a}inen and Karl Renk for discussions.
This research was partially supported by the Academy of Finland (grant 163358)
and NorFA.

%
\vspace{3cm}
\begin{figure}
\epsfxsize=7cm
\hspace{3cm}
\epsfbox{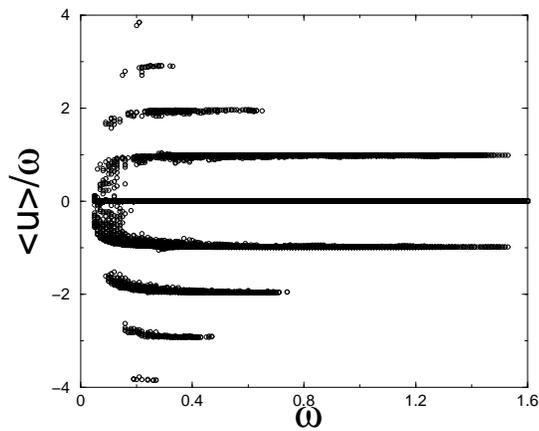}
\caption{The dependence of spontaneously generated dc bias $\langle
u\rangle/\omega$
on ac frequency $\omega$, scaled to the miniband plasma frequency
$\omega_{pl}$, and for $\gamma_v=0.1\omega_{pl}$,
$\gamma_\varepsilon=\alpha=0.01\omega_{pl}$.}
\end{figure}

\vspace{0.5cm}
\begin{figure}
\epsfxsize=7cm
\hspace{3cm}
\epsfbox{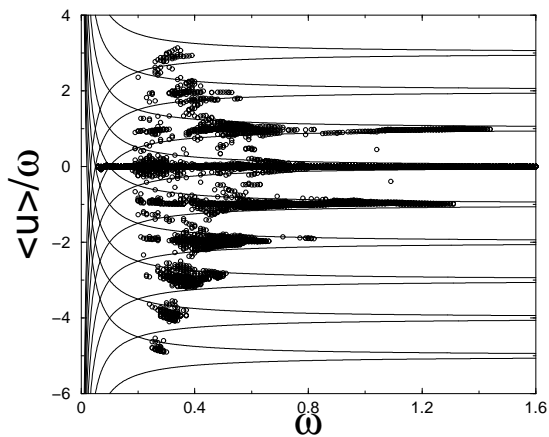}
\caption{Same as in Fig.1, but for
$\gamma_v=\gamma_\varepsilon=0.1\omega_{pl}$,
$\alpha=0.01\omega_{pl}$.}
\end{figure}

\vspace{0.5cm}
\begin{figure}
\epsfxsize=7cm
\hspace{3cm}
\epsfbox{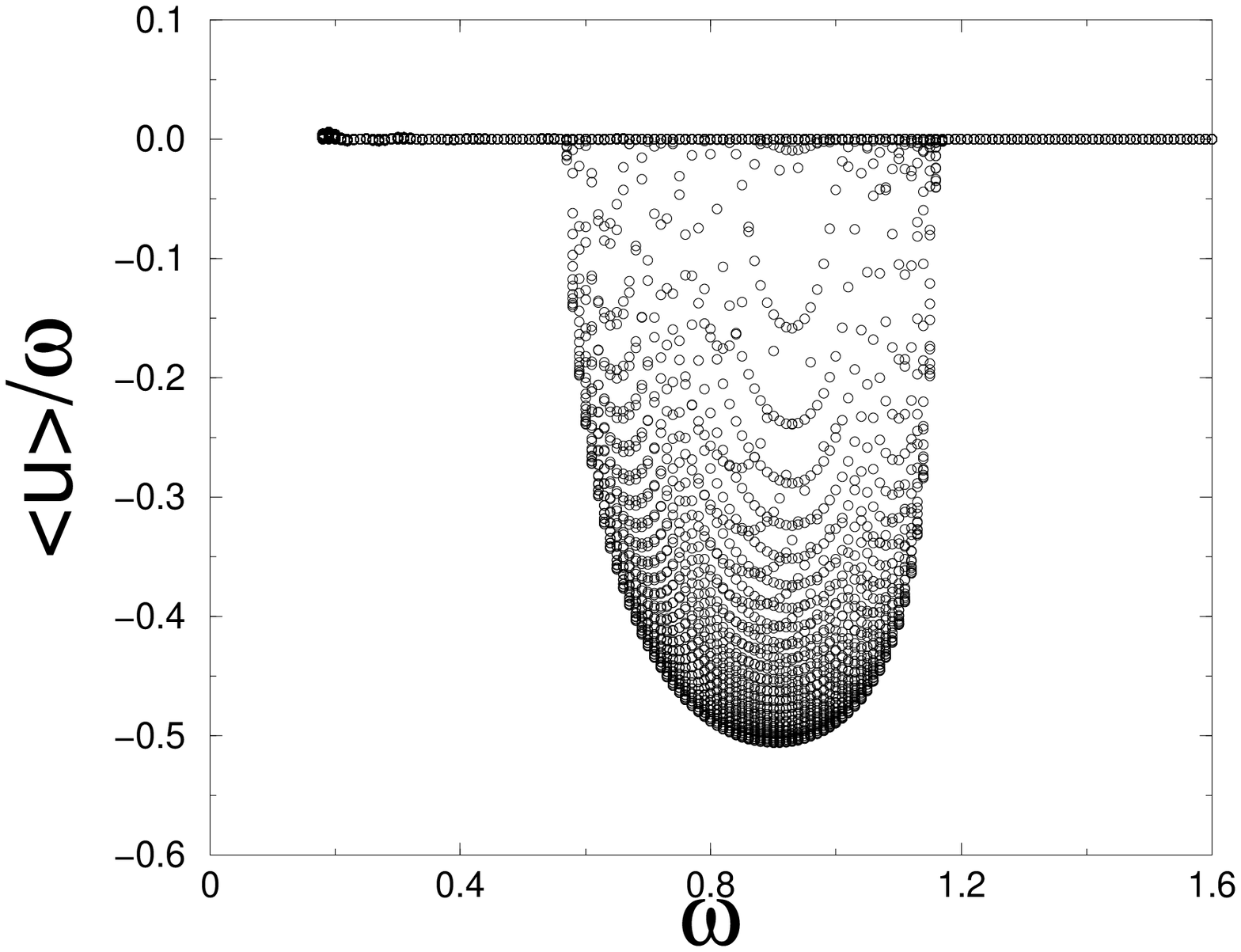}
\caption{Same as in Figs.1 and 2, but for strong
damping: $\gamma_v=\omega_{pl}$, $\gamma_\varepsilon=0.1\omega_{pl}$,
$\alpha=0.03\omega_{pl}$.}
\end{figure}


\begin{thebibliography}{99}
\bibitem{bass}
For a review of the relevant early work, see
Bass F. G. and Tetervov A. P., Phys. Rep., {\bf 140} (1986) 237.
\bibitem{exp-thz}
Unterrainer K.,{\it et al.}, Phys. Rev. Lett., {\bf 76} (1996) 2973;
Winnerl S. {\it et al.}, Appl. Phys. Lett., {\bf 77} (2000) 1259.
\bibitem{winnerl97}
Winnerl S. {\it et al.}, Phys. Rev. B, {\bf 56} (1997) 10303.
\bibitem{exp-samples}
Stormer H. L. {\it et al.}, Appl. Phys. Lett., {\bf 58} (1991) 726;
Majumdar A. {\it et al.}, Appl. Phys. Lett., {\bf 76} (2000) 3600.
\bibitem{multistab}
Ghosh A. W., Kuznetsov A. V. and Wilkins J. W., Phys. Rev. Lett.,
{\bf 79} (1997) 3494; Dodin E. P., Zharov A. A. and  Ignatov A. A.,
Zh. Eksp. Teor. Fiz., {\bf 114} (1998) 2246 [JETP, {\bf 87} (1998)
1226];
Ghosh A. W. {\it et al.}, Appl. Phys. Lett., {\bf 74} (1999) 2164;
Romanov Yu. A. and Romanova Yu. Yu., Fiz. Tekh. Polupr., {\bf 35}
(2001) 211
[Semiconductors, {\bf 35} (2001) 204].
\bibitem{zharov00}
Zharov A. A., Dodin E. P. and Raspopin A. S.,
Pis'ma Zh. Eksp. Teor. Fiz., {\bf 72} (2000) 653  [JETP Lett., {\bf
72}
(2000) 453].
\bibitem{alekseev96}
Alekseev K. N. {\it et al.}, Phys. Rev. B, {\bf 54} (1996) 10625.
\bibitem{cao00}
Cao J. C., Liu H. C. and Lei X. L., Phys. Rev. B, {\bf 61} (2000)
5546.
\bibitem{yevtushenko}
Yevtushenko O. M. and Richter K., Phys. Rev. B, {\bf 57} (1998) 14839;
Physica E, {\bf 4} (1999) 256.
\bibitem{ignatov95}
Ignatov A. A. {\it et al.}, Z. Phys. B, {\bf 98} (1995) 187.
\bibitem{alekseev98}
Alekseev K. N. {\it et al.}, Phys. Rev. Lett., {\bf 80} (1998) 2669;
Physica D, {\bf 113} (1998) 129.
\bibitem{alekseev99}
Alekseev K. N., Erementchouk  M. V. and Kusmartsev F. V.,
Europhys. Lett., {\bf 47} (1999) 595.
\bibitem{romanov00}
Romanov Yu. A. and Romanova Yu. Yu., Zh. Eksp. Teor. Fiz., {\bf 118}
(2000) 1193 [JETP, {\bf 91} (2000) 1033].
\bibitem{anc}
Ignatov A. A. and Romanov Yu. A., Fiz. Tverd. Tela, {\bf 17} (1975)
3388
[Sov. Phys. Solid State, {\bf 17} (1975) 2216];
Phys. Stat. Sol. (b), {\bf 73} (1976) 327.
\bibitem{dunlap86}
Dunlap D.H. and Kenkre V. M., Phys. Rev. B, {\bf 34} (1986) 3625.
\bibitem{holthaus}
Holthaus M., Phys. Rev. Lett., {\bf 69} (1992) 351.
\bibitem{dunlap93}
Dunlap D. H. {\it et al.}, Phys. Rev. B, {\bf 48} (1993) 7975.
\bibitem{wave-mix}
Esaki L. and Tsu R., Appl. Phys. Lett., {\bf 19} (1971) 246;
Romanov Yu. A., Opt. Spektr.,  {\bf 33} (1972) 917
[ Sov. Phys. Opt. Spectr., {\bf 33} (1972) 503  ].

\bibitem{modulation-doping}
Dingle R. {\it et al.}, Appl. Phys. Lett., {\bf 19} (1971) 246.
\bibitem{alekseev00}
Alekseev K. N., 2001 (unpublished).

\bibitem{kautz}
Kautz R. L., Rep. Prog. Phys., {\bf 59} (1996) 935.

\bibitem{d'humieres}
D'Humieres D. {\it et al.}, Phys. Rev. A {\bf 26} (1982) 3483.

\bibitem{sb-def}
Levinsen M. T., J. Appl. Phys. {\bf 53} (1982) 4294;
Miles J., Physica D  {\bf 31} (1988) 252.

\bibitem{sb-strong}
McDonald A. H. and Plischke M., Phys. Rev. B {\bf 27} (1983) 201;
Octavio M., Phys. Rev. B {\bf 29} (1984) 1231.

\bibitem{helm}
Helm M. {\it et al.}, Phys. Rev. Lett., {\bf 82} (1999) 3120.

\bibitem{vagov}
Kusmartsev F. V. and Vagov A.V., 2001 (unpublished).
\bibitem{thesis}
Cannon E. H., Ph. D. thesis, University of Illinois at
Urbana-Champaign,
1999.

\end{thebibliography}
\end{document}